# Complex networks: A mixture of power-law and Weibull distributions

Ke Xu[*], Liandong Liu, Xiao Liang

State Key Laboratory of Software Development Environment
Beihang University, Beijing 100191, China

**Abstract:** Complex networks have recently aroused a lot of interest. However, network edges are considered to be the same in almost all these studies. In this paper, we present a simple classification method, which divides the edges of undirected, unweighted networks into two types: p2c and p2p. The p2c edge represents a hierarchical relationship between two nodes, while the p2p edge represents an equal relationship between two nodes. It is surprising and unexpected that for many real-world networks from a wide variety of domains (including computer science, transportation, biology, engineering and social science etc), the p2c degree distribution follows a power law more strictly than the total degree distribution, while the p2p degree distribution follows the Weibull distribution very well. Thus, the total degree distribution can be seen as a mixture of power-law and Weibull distributions. More surprisingly, it is found that in many cases, the total degree distribution can be better described by the Weibull distribution, rather than a power law as previously suggested. By comparing two topology models, we think that the origin of the Weibull distribution in complex networks might be a mixture of both preferential and random attachments when networks evolve.

Complex networks have attracted great interest in the past decade [1-2] and have been extensively studied in various contexts, such as the Internet whose nodes are Autonomous Systems and edges represent traffic exchange relationships [2,3,15], protein-protein interaction networks whose nodes are proteins and edges represent physical interactions [5], transport networks whose nodes are cities and edges represent airlines, and social networks whose nodes are individuals and edges represent friendships or partnerships [7-9]. It is one of the most well-known discoveries in the study of complex networks that such networks have a scale-free characteristic [2] which is that the degree distribution $P(k)$ follows a power law (i.e., $P(k) \sim k^{-r}$), where $P(k)$ is the fraction of nodes having $k$ connections to other nodes. This discovery improves our understanding of network structure and evolution, and has a large impact upon the various applications of a network.

However, there has been some debate about the power-law distribution [10-14] recently. For example, some researchers have shown that the degree distribution does not strictly follow a power law [11-14]. In addition, when the networks are investigated, the edge type is mostly ignored and all edges are considered equal. For instance, according to different business relationships, edges of the Internet at the Autonomous System (AS) level fall into two categorical types: provider-to-customer and peer-to-peer. He et al. [3] calculated the degree distribution according to the edge type. They found that the provider-to-customer degree distribution can be well-fitted by a power law while the peer-to-peer degree distribution can be well-fitted by a Weibull distribution. The total degree distribution can be considered as a mixture of the power-law and Weibull distribution (see Figure 1). Motivated by this finding, we propose that there are two types of edges in complex networks. One of them follows a power-law degree distribution while the other one follows another degree distribution (e.g., the Weibull distribution). When the majority of the network edges are of the former type, its total degree distribution follows the power law. But as the number of latter edges increases, the total degree distribution does not follow the power law as strictly as before. If this conjecture can be verified, it may help to progress the debate about the power-law distribution of complex networks.

---

[*] To whom correspondence should be addressed. Email: kexu@nlsde.buaa.edu.cn.

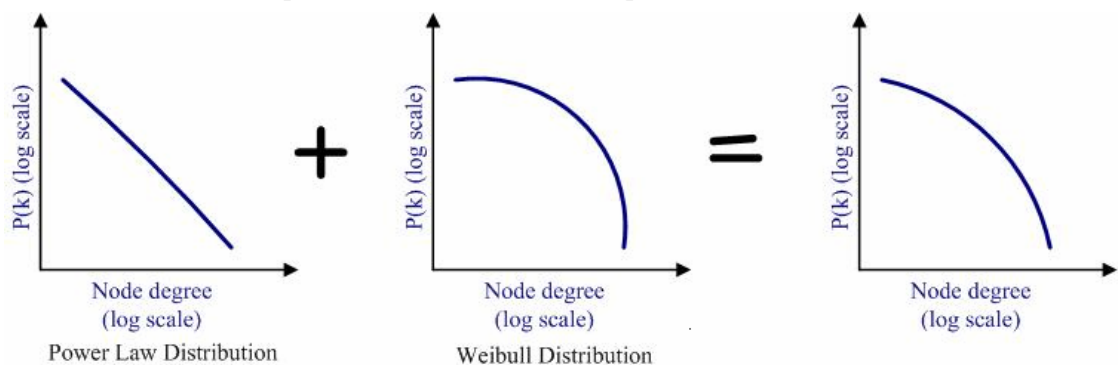

Figure 1. A mixed degree distribution of a power-law and Weibull distribution

To confirm our conjecture, we investigated whether the same phenomenon that exists in the Internet is found in other complex networks. But for other networks, there is no uniform classification of edges, so a universal method of edge classification must be designed first. From the definition of provider-to-customer and peer-to-peer relationships on the Internet, we see that a peer-to-peer edge always connects two nodes at the same level while a provider-to-customer edge always connects two nodes at different levels. On account of this, we propose a universal method of edge classification based on the levels of networks. The details are described below.

For an undirected and unweighted graph $G$, its edges will be classified into two types after three steps. First, we identify the center nodes in $G$. In this paper, the center nodes are defined as the nodes having the minimum eccentricity among all the nodes, where the eccentricity of a node is the greatest distance between it and any other node. Then we determine the level of nodes. All the center nodes are at the first (highest) level and the level of a node is higher as it's closer to the center nodes. For example, if a node can reach a center node through two hops (edges) and this is the shortest distance between this node and any center node, then we say that it is at the third level. Finally, we classify an edge into one of the two types according to the level of its two endpoints. If the two endpoints of an edge come from the same level, then it is considered a peer-to-peer edge. Otherwise, it is considered a provider-to-customer edge. For convenience, provider-to-customer and peer-to-peer are called p2c and p2p respectively in the rest of this paper.

From the discussion above, we see that the p2c and p2p edges have different functions. If the connectivity of the center nodes is guaranteed, then the whole network can be connected through only the p2c edges. But the p2p edges can reduce the distance among nodes, enhance robustness, and improve the performance and efficiency of the whole network. If a network is compared with a human body, the p2c edges can be imagined as the bones while the p2p edges can be imagined as the muscles. The p2c edges support the whole network but the p2p edges make the network more energetic.

**Applications to the Internet AS-level topology**

To validate our classification method, we applied it to the two Internet AS-level topologies collected by Zhang et al [4] and He et al [3]. For the former, 149 nodes were classified as center nodes from a total of 30850 nodes. The 106375 edges in total were divided into 63528 p2c edges and 42847 p2p edges. The types of edges classified by our method are the same as the actual types in 68.7% of cases. For the latter, 7 nodes were classified as center nodes from a total of 19936 nodes. The 59508 edges in total were divided into 34733 p2c edges and 24775 p2p edges. The types of edges classified by our method are the same as the actual types in 75.2% of cases. We can see that our classification is not exactly the same as the one in Internet AS-level topologies, but there is reasonable consistency in edge types. This shows that they represent some kind of relationship between two nodes. P2p edges represent a kind of equal relationship while p2c edges represent a kind of unequal or hierarchical relationship. We respectively plotted the complementary cumulative distribution functions (CCDF) of the total, p2c and p2p degree distribution on a log-log scale in Figure 2. For both of the two Internet AS-level topologies, the p2c degree distribution follows a power-law distribution more strictly than the total degree distribution, while the p2p degree distribution follows the Weibull distribution very well. All the correlation coefficients are higher than 97% (see Table 1). These results validate our classification method.

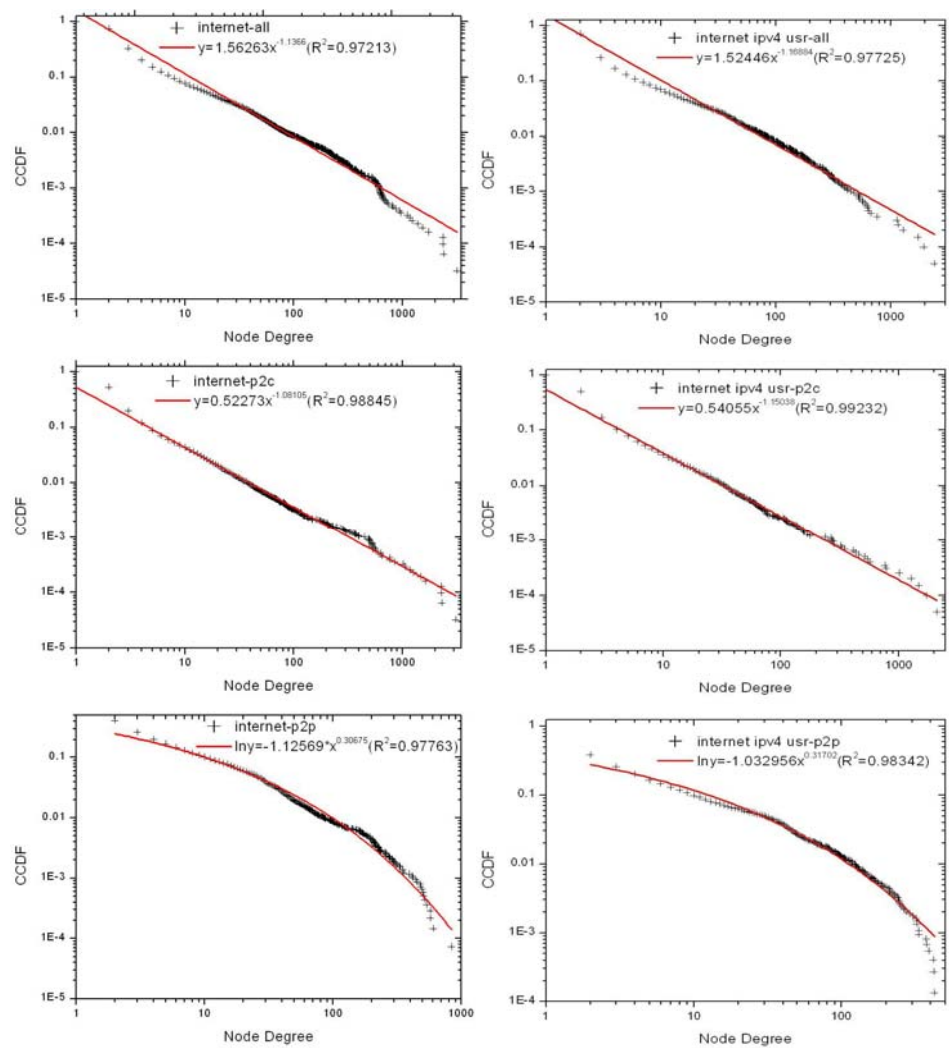

Figure 2. The total degree distributions of the Internet AS level topology collected by Zhang et al (left) and He et al

(right) in the top row, their p2c degree distribution in the middle row and their p2p degree distribution la the bottom row.

Table 1. Summary of the results that our classification produces on ten real complex networks, including the number of nodes (N), the number of edges (E), the number of p2c edges (P2C), the number of p2p edges (P2P), the correlation coefficient of fitting a power law to the total degree distribution ($R_{PL}^t$), the correlation coefficient of fitting a power law to the p2c degree distribution ($R_{PL}^{p2c}$), the correlation coefficient of fitting a Weibull distribution to the p2p degree distribution ($R_W^{p2p}$) and the correlation coefficient of fitting a Weibull distribution to the total degree distribution ($R_W^t$).

|  | N | E | P2C | P2P | $R_{PL}^t$ (%) | $R_{PL}^{p2c}$ (%) | $R_W^{p2p}$ (%) | $R_W^t$ (%) |
|---|---|---|---|---|---|---|---|---|
| IPv4 (Zhang et al) | 30850 | 106375 | 63528 | 42847 | 97.20 | **98.85** | 97.76 | 88.09 |
| IPv4 (He et al) | 19936 | 59508 | 34733 | 24775 | 97.73 | **99.23** | 98.34 | 89.40 |
| IPv6 | 419 | 1812 | 1025 | 787 | 89.59 | **95.58** | 97.30 | 94.21 |
| US Airline | 233 | 1612 | 714 | 898 | 82.63 | **96.17** | 98.26 | **98.24** |
| Yeast | 2224 | 6609 | 3882 | 2727 | 89.08 | 94.56 | 98.81 | **99.26** |
| Power Grid | 4941 | 6594 | 5660 | 934 | 88.82 | 89.91 | **99.20** | 95.09 |
| FaceBook | 63392 | 816886 | 343067 | 473819 | 87.01 | **91.90** | 99.64 | 99.20 |
| Email | 1133 | 5451 | 2677 | 2774 | 77.76 | 86.44 | **99.62** | **99.74** |
| NetScience | 379 | 914 | 461 | 453 | 90.62 | **97.28** | 90.74 | 90.11 |
| Geom | 3621 | 9461 | 5295 | 4166 | 92.97 | **96.13** | 97.30 | 91.59 |

**Applications to eight other networks**

To demonstrate the universality of the phenomenon shown in this paper, we applied our classification method to eight other real-world networks. The eight undirected, unweighted networks we chose were drawn from a variety of different domains, including computer science, transportation, biology, engineering and social science etc, which were as follows.

a) IPv6: The IPv6 Internet topology at the AS level collected by dolphin [15].
b) US Airline: the U.S airline network in 1997
c) Yeast: the protein-protein interaction network in budding yeast [5]
d) Power Grid: the network of the Western States Power Grid of the United States [6]
e) Facebook (New Orleans): the New Orleans regional network in Facebook , whose nodes are users and edges represent friendships [7]
f) Email: the network of e-mail interchanges between members of the University Rovira i Virgili (Tarragona)
g) NetScience: a coauthorship network of scientists working on network theory and experiment [9]
h) Geom: a coauthorship network of scientists working on computational geometry [8]

As shown in Table 1, the p2c degree distribution follows a power law more strictly than the total degree distribution. For all networks except the Power Grid and Email, the correlation coefficients are more than 90% and most of them are over 95%. Surprisingly, the p2p degree distribution of all the eight networks follows the Weibull distribution very well and the correlation coefficients are over 97%, excepting NetScience. These results strongly suggest that it is universal that the total degree distribution of complex networks can be considered as a mixture of power-law and Weibull distributions. More surprisingly, we can see that for six networks (e.g. IPv6), the total degree distribution is closer to the Weibull distribution, rather than a power law as previously suggested.

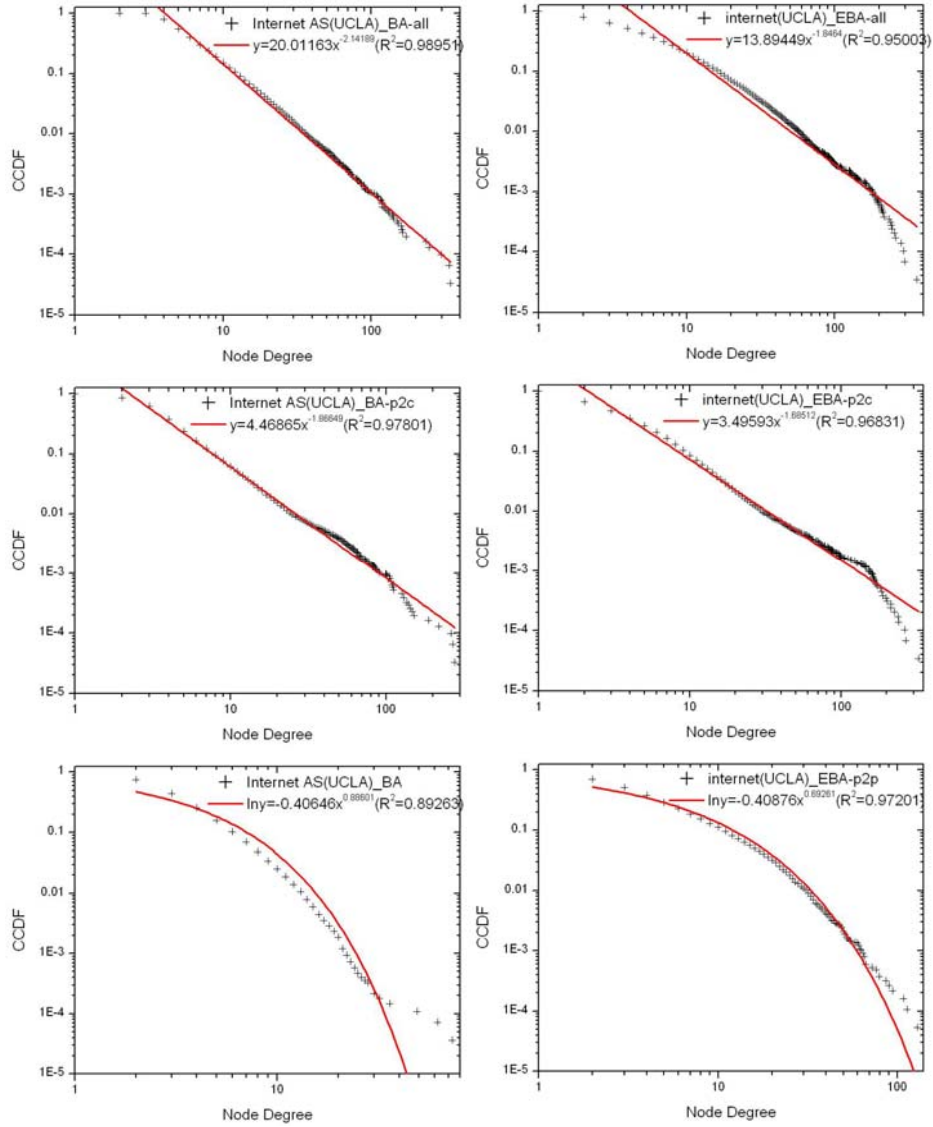

Figure3 The total, p2c and p2p degree distribution of the network constructed by BA and EBA models. (a) In the simulations with the BA model, the parameters we used were: m=3, $m_0$=2 and t=30848. (b) In the simulations with the EBA model, the parameters we used were: p=0.35, q =0.5, m=1, $m_0$=2 and t=213680.

Table 2. Comparison of the results that our classification produces on the network generated by BA and EBA models.

|  | N | E | P2C | P2P | $R_{PL}^t$ (%) | $R_{PL}^{p2c}$ (%) | $R_W^{p2p}$ (%) |
|---|---|---|---|---|---|---|---|
| source | 30850 | 106375 | 63528 | 42847 | 97.20 | 98.85 | 97.76 |
| BA | 30850 | 106546 | 65447 | 41099 | 98.95 | 97.80 | **89.27** |
| EBA | 29166 | 106318 | 63644 | 42674 | 95.00 | 96.83 | **97.20** |

**The origin of the Weibull distribution**

The power-law degree distribution can be seen as the result of rich-get-richer behavior: new nodes attach preferentially to high degree nodes. Based on this, Barabasi and Albert proposed the well-known BA model [2], and then extended it through the addition of new edges, and the rewiring of edges [1]. In this paper, we call the extended version of the BA model the EBA model. In order to explain the origin of the phenomenon found in this paper, we respectively simulated the Internet AS-level topology with the BA and EBA model, and then applied our classification to the networks we constructed. It is surprising to us that there is the same phenomenon (the total degree distribution is a mixture of the power-law and Weibull distribution) in the network constructed by the EBA model as in the real Internet, but there is no such phenomenon in the network constructed by the BA model (see Figure 3 and Table 2). Comparing the EBA model with the BA model, we think that the phenomenon may be caused by the process of adding and rewiring edges when networks evolve. The p2p edges follow the Weibull degree distribution, which is different from a power law, because a node is selected as an endpoint of a new edge with not only a degree-preferential probability but also a random probability when adding or rewiring an edge.